\documentclass[prl,showpacs,byrevtex,floatfix,twocolumn]{revtex4}
\usepackage{graphicx}
\usepackage{dcolumn}
\usepackage{bm}

\begin{document}

\bibliographystyle{apsrev}

%\preprint{Draft - not for distribution}

\title[hfm]{Heavy fermion fluid in high magnetic fields: an
infrared study of CeRu$_4$Sb$_{12}$}

\author{S. V. Dordevic$^{1, \dag}$}
\email{sasha@physics.uakron.edu}%
\author{K.S.D. Beach$^2$}
\author{N. Takeda$^3$}
\author{Y.J. Wang$^4$}
\author{M.B. Maple$^1$}
\author{D.N. Basov$^1$}

\affiliation{$^1$ Department of Physics, University of California,
San Diego, La Jolla, CA 92093}

\affiliation{$^2$ Department of Physics, Boston University,
Boston, MA 02215}

\affiliation{$^3$ Institute for Solid State Physics, University of
Tokyo, 5-1-5 Kashiwa, Chiba 277-8581, Japan}

\affiliation{$^4$ National High Magnetic Field Laboratory,
Tallahassee, FL 32310}

\date{\today}

%
% The abstract goes here
%
\begin{abstract}
We report a comprehensive infrared magneto-spectroscopy study of
CeRu$_4$Sb$_{12}$ compound revealing quasiparticles with heavy
effective mass m$^*$, with a detailed analysis of optical
constants in fields up to 17\,T. We find that the applied magnetic
field strongly affects the low energy excitations in the system.
In particular, the magnitude of m$^*$\,$\simeq$\,70\,m$_b$ (m$_b$
is the quasiparticle band mass) at 10\,K is suppressed by as much
as 25\,$\%$ at 17 T. This effect is in quantitative agreement with
the mean-field solution of the periodic Anderson model augmented
with a Zeeman term.
\end{abstract}

 %
% PACS numbers
%
% 74.25.-q - General properties; correlations between normal and
% SC states
% 74.25.Gz - optical properties of superconductors
% 78.30.-j - Infrared and Raman spectra
%


\maketitle

\makeatletter%
\global\@specialpagefalse%
\let\@evenhead\@oddhead%
\makeatother

Systematic investigations of physical properties of Heavy Fermion
(HF) materials have provided valuable insights into the role
played by many body effects in transport and magnetic phenomena
\cite{hewson97}. The HF ground state is believed to be a
manifestation of a delicate balance between competing interactions
\cite{doniach77}. External stimuli can readily tip the balance
yielding a variety of novel forms of matter. Motivated by the
exceptional richness and complexity of effects occurring in the HF
systems under applied external stimuli, many research teams have
undertaken the methodical investigation of properties under the
variation of relevant control parameters, such as pressure (P) or
magnetic field (H) \cite{gegenwart03,kim03,harrison03}. In this
work we focused on the analysis of the low energy excitations of
the HF system CeRu$_{4}$Sb$_{12}$ in high magnetic field using
infrared (IR) spectroscopy. This technique has emerged as a very
powerful experimental probe of the HF ground state
\cite{degiorgi99,dordevic01}. Essentially all hallmark features of
the the HF state, such as the quasiparticle effective mass
m$^{\ast }$, the optical gap $\Delta $, and a narrow Drude--like
peak (with the width 1/$\tau $) can be simultaneously studied
based on the analysis of the optical constants.

Our IR magneto-optics results uncover robustness of the HF fluid
in CeRu$_{4}$Sb$_{12}$ against magnetic field: a field of 17\,T
suppresses the quasiparticle effective mass by about 25\,$\%$. The
reason we selected CeRu$_{4}$Sb$_{12}$ for high-field
investigations reported here is two-fold. First, the family of
skutterudite compounds to which CeRu$_{4}$Sb$_{12}$ belongs
reveals a broad spectrum of behavior, including non-Fermi liquid
power laws, superconductivity with unconventional order parameter
and antiferromagnetism \cite{takeda99,takeda00,bauer01,bauer02}.
This catalogue of enigmatic properties supports the notion of the
competing ground states: a regime where even modest external
stimuli are likely to have a major impact on the response of a
system. Second, our earlier zero-field experiments have uncovered
remarkably large changes of reflectance R($\omega $) of
CeRu$_{4}$Sb$_{12}$ at low temperatures: the reflectance drops by
more than 20$\%$ in the far-IR \cite{dordevic01}. With such a
strong feature, unprecedented among HF compounds, a task of
monitoring subtle field-induced modifications of electrodynamics
is becoming a viable endeavor.

Samples of CeRu$_{4}$Sb$_{12}$ were synthesized using Sb--flux
method with an excess of Sb (for more details on sample growth see
Ref.~\onlinecite{takeda99}). Large polycrystalline specimens with
a typical surface area of 4$\times $4\thinspace mm$^{2}$ were
essential for magneto--optical experiments. The samples have been
characterized by resistivity and magnetization measurements and
found to be of similar quality as previously reported
\cite{takeda99}.

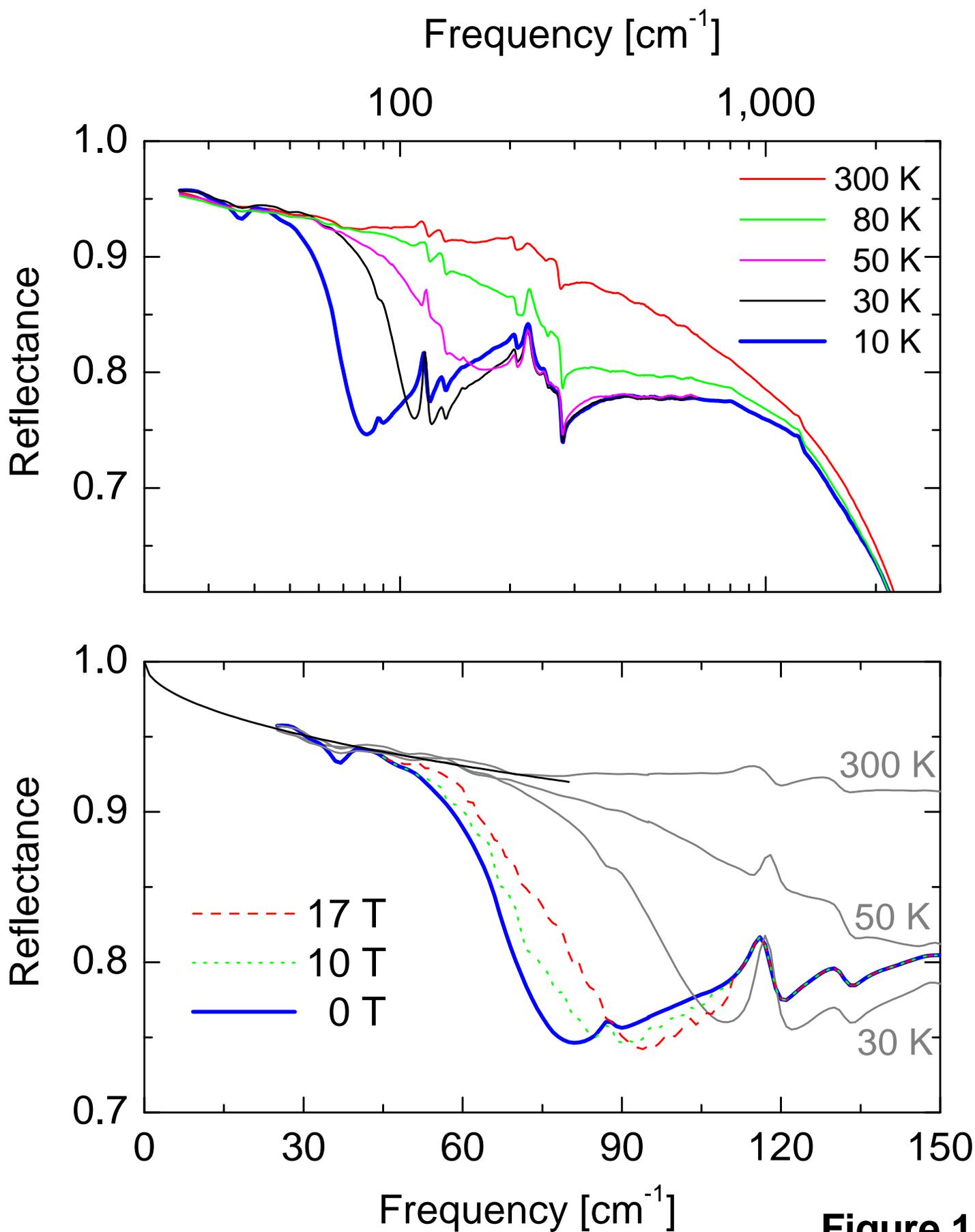
\begin{figure}%[tbp]
\caption{Temperature and field dependence of the reflectance of
CeRu$_4$Sb$_{12}$. Top panel: temperature dependence with H=\,0.
Bottom panel: high field reflectance at T=10\,K, along with
several zero field curves at higher temperatures.} \label{fig:ref}
\end{figure}

The electrodynamic response of CeRu$_{4}$Sb$_{12}$ was studied
using infrared and optical reflectance spectroscopy. Zero field
near--normal--incidence reflectance R($\omega $) was measured in
the frequency range 30--30,000\thinspace cm$^{-1}$ (approximately
4\thinspace meV--4\thinspace eV) from T=10\thinspace K to room
temperature. These measurements were supplemented with
magneto--optic reflectance ratios R($\omega $,\thinspace
H)/R($\omega $,H=0\thinspace T) obtained in zero--field cooling
conditions at 10\thinspace K \cite{padilla04}. The absolute value
of reflectance in magnetic field R($\omega $,\thinspace H) was
obtained by multiplying these ratios with the H=0 data R($\omega
$,\thinspace H)=[R($\omega $,\thinspace H)/R($\omega
$,H=0\thinspace T)]$\cdot $R($\omega $). The complex optical
conductivity $\sigma (\omega )=\sigma _{1}(\omega )+i\sigma
_{2}(\omega )$ and dielectric function $\epsilon (\omega
)=\epsilon _{1}(\omega )+i\epsilon _{2}(\omega )$ were obtained
from reflectance using Kramers--Kronig analysis. For low-frequency
extrapolations, we used the Hagen-Rubens formula (thin line in
Fig.~\ref{fig:ref}). The overall uncertainty in $\sigma (\omega )$
and $\epsilon (\omega )$ due to low- and high-frequency
extrapolations required for Kramers--Kronig analysis is about 8-10
$\%$. This uncertainty does not affect any conclusions of the
paper in any significant way.

Plotted in the top panel of Fig.~\ref{fig:ref} is the temperature
dependence of reflectance R($\omega $) over an extended frequency
interval. The gross features of R($\omega $) are in good agreement
with the spectra obtained previously on single crystals
\cite{dordevic01}. The general shape of the reflectance spectrum
at 300\thinspace K is metallic, with high absolute values at low
frequencies, and a well defined edge around 4,000\thinspace
cm$^{-1}$ (not shown). Below a coherence temperature $T^{\ast }$
which is about 50 K for CeRu$_{4}$Sb$_{12}$ \cite{dordevic01}
there is a strong suppression of reflectance in the range between
50 and 1,000 cm$^{-1}$. This result is indicative of the
development of a gap in the density of states commonly observed in
HF systems. Simultaneously a new feature develops in the far-IR.
This structure bears similarities with the conventional "plasma
edge" in metals and is established as a spectroscopic signature of
heavy quasiparticles \cite{degiorgi99}. The reflectance edge
softens with decrease of T down to 60 cm$^{-1}$ in the 10 K
spectra. Several narrow peaks in the far--IR region are due to
optically active phonons and will be discussed below.

The bottom panel of Fig.~\ref{fig:ref} displays the reflectance
data in magnetic field, plotted on a linear scale over a narrower
frequency interval in order to emphasize field-induced changes.
Only a few in-field curves are shown for clarity. For comparison,
the plot also includes several zero field spectra at higher
temperatures. The thin line is a Hagen--Rubens extrapolation used
for the 300\thinspace K spectrum. For fields below 5\thinspace T
no significant change in reflectance can be detected, within the
error bars of the experiment. At higher fields, the frequency
position of the low lying "plasma edge" hardens, a behavior which
is paralleled by zero-field spectra taken at higher T. Comparing
temperature- and field-induced modification of R($\omega $), we
notice that the net effect of applying a 17 T field is
functionally equivalent to a temperature increase of $\approx
$\thinspace 20\thinspace K. At higher frequencies ($\omega \gtrsim
$\thinspace 110\thinspace cm$^{-1}$) the reflectance is field
independent for H\thinspace $<$\thinspace 17\thinspace T, within
the error bars of the present experiment.

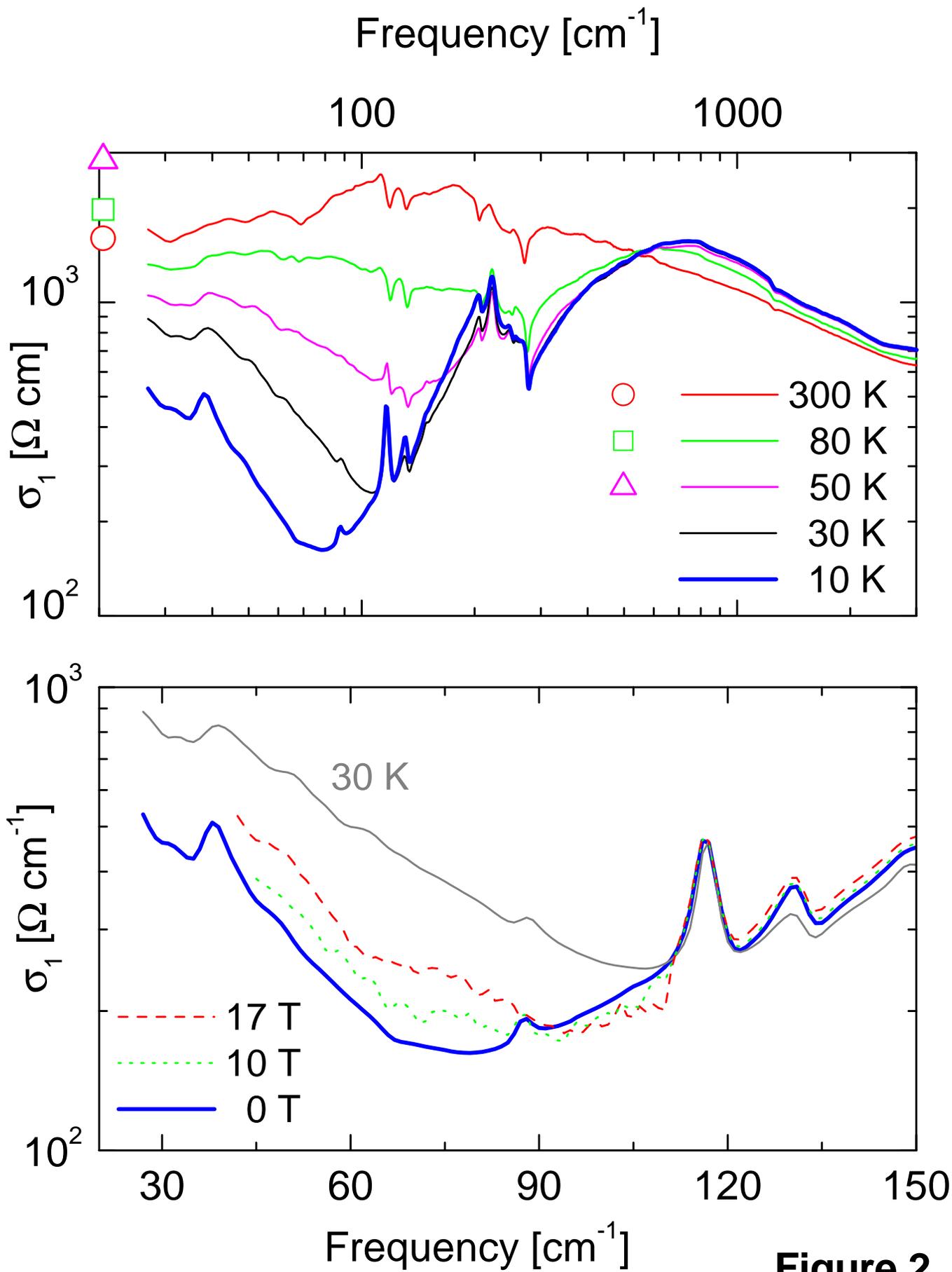
\begin{figure}%[tbp]
\caption{Temperature and field dependence of the real part of the
optical conductivity $\protect\sigma _1(\protect\omega)$. Top
panel: temperature dependence in zero field. Bottom panel: optical
conductivity in high magnetic field at T=10\,K, along with several
zero field curves at higher temperatures.} \label{fig:cond}
\end{figure}

The top panel of Fig.~\ref{fig:cond} displays the temperature
dependence of the dissipative part of the optical conductivity
$\sigma _{1}(\omega )$. The dominant feature in the low-T data is
a sharp increase in the conductivity above 80\thinspace cm$^{-1}$
with an overshoot of the 300 K spectrum at 550\thinspace
cm$^{-1}$. This behavior is usually assigned to the formation of
the (hybridization) gap in the excitation spectrum of a HF system
\cite{degiorgi99}. Because of the improved signal-to-noise ratio,
it became possible to extend earlier zero field measurements
\cite{dordevic01} to lower frequencies. These new results allow us
to resolve the onset of a narrow Drude-like feature, usually
associated with the response of heavy quasiparticles at low T
\cite{degiorgi99}. The gradual suppression of the conductivity at
$\omega <$\thinspace 60\thinspace cm$^{-1}$ at low T is consistent
with the decrease of the quasiparticle scattering rate 1/$\tau
(\omega )$ below the low-$\omega $ cut-off of our data (30
cm$^{-1}$). The latter effect will lead to an increase of the
conductivity in the $\omega \rightarrow $\thinspace 0 limit, in
accord with the DC resistivity measurements
\cite{takeda00,bauer01,abe02} displayed with full symbols. The
narrow resonances at 88, 116, 131, 205, 221 and 248 cm$^{-1}$ are
phonon modes. Some of these modes, including the peaks at 88, 131
and 205\thinspace cm$^{-1}$, have not been observed before in
measurements on smaller single crystals \cite{dordevic01}. We also
observe an additional mode at 40\thinspace cm$^{-1}$ which can be
identified only in the low-T spectra once the electronic
background in the conductivity is diminished.

The bottom panel of Fig.~\ref{fig:cond} displays the optical
conductivity $\sigma _{1}(\omega )$ at 10 K in a magnetic field,
along with 30\thinspace K data (in zero field). Only a limited
frequency region is shown to emphasize the effects due to applied
fields. We find that the onset of the gap structure in
$\sigma_1(\omega)$ is virtually unaffected even by the
17\thinspace T field. On the other hand, the coherent mode due to
heavy quasiparticles broadens as the field increases. This is
consistent with positive magneto-resistance in the DC data
\cite{takeda00,bauer01,abe02}.

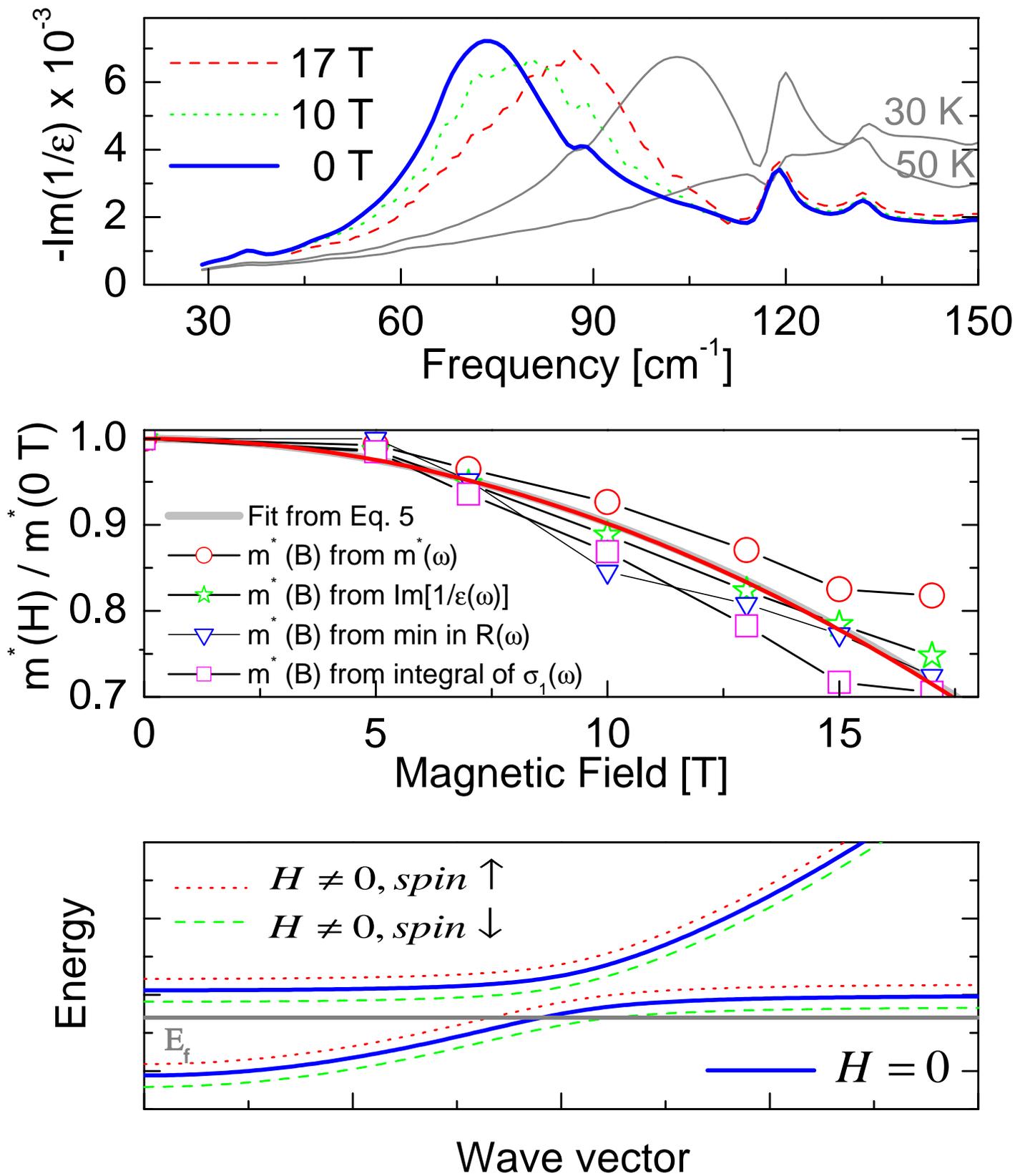
\begin{figure}%[tbp]
\caption{The top panel presents the loss function
Im[1/$\protect\epsilon(\protect\omega)$], which has pronounced
peaks at the frequencies of the longitudinal modes, such as the
heavy plasmon $\protect\omega_p^*$. All field curves are at
T=10\,K. The middle panel shows the field dependence of
quasiparticle effective mass m$^*$ extracted from: 1) the
m$^*(\protect\omega)$ spectra, 2) the maximum in the loss function
Im[1/$\protect\epsilon(\protect\omega)$], 3) minimum in
reflectance R($\omega$) and 4) the integral of the narrow Drude
mode (Eq.~\ref{eq:sum}). The bottom panel schematically displays
the PAM band structure, both in zero and high magnetic field.}
\label{fig:scaling}
\end{figure}

Field induced effects in the optical constants can be quantified
using several complementary methods. The extended Drude model
(EDM) offers an instructive analysis protocol and is routinely
used to quantify correlation effects in HF systems
\cite{degiorgi99}. Within the EDM the optical constants are
expressed in terms of the quasiparticle scattering rate $1/\tau
(\omega )$ and the effective mass spectrum $m^{\ast }(\omega )$
\cite{degiorgi99}. Specifically, $m^{\ast }(\omega )$ can be
obtained from the following equation:

\begin{equation}
\frac{m^{*}(\omega)}{m_b} = - \frac{\omega_{p}^{2}}{4 \pi \omega}
Im \Big[\frac{1} {\sigma(\omega)}\Big],  \label{eq:mass}
\end{equation}
where the plasma frequency $\omega_p$\,=\,$\sqrt{4 \pi e^2
n/m_b}$\,=\,10,744\,cm$^{-1}$ is estimated from the integration of
$\sigma_{1}(\omega)$ up to the frequency of the onset of interband
absorption. The spectra of $m^*(\omega)$ in HF compounds usually
reveal a complicated frequency dependence which is particularly
true in the vicinity of the energy gap \cite{degiorgi99}. However,
an extrapolation of $m^{*}(\omega)$ spectra towards zero frequency
yields a reliable estimate of the quasiparticle effective mass
m$^{\ast}$ that has been shown to agree with the thermodynamic
mass for a variety of HF compounds \cite{degiorgi99,dordevic01}.
The estimated value of $\simeq$\,70\,m$_b$ (at 10\,K)
CeRu$_4$Sb$_{12}$ is somewhat smaller then previously reported
80\,m$_b$ for single crystals \cite{dordevic01}. This effect can
be assigned to increased disorder and/or variations of the Kondo
temperature in polycrystalline specimens compared with previously
studied single crystals. In the middle panel of
Fig.~\ref{fig:scaling} we plot the extracted values of the mass in
zero frequency limit as a function of magnetic field. A 17\,T
field is capable of suppressing the mass by about 25\,$\%$ from
its zero-field value.

Alternatively, mass renormalization can be inferred from the
analysis of the loss function spectra Im[1/$\epsilon (\omega )$]
(top panel of Fig.~\ref{fig:scaling}). The loss function offers a
convenient way to explore longitudinal modes in the optical
spectra, including heavy electron plasmons in HF compounds
\cite{degiorgi99}. These longitudinal modes produce
Lorentzian-like peaks in Im[1/$\epsilon (\omega )$] spectra
centered at $\omega _{p}^{\ast }$\thinspace =\thinspace
$\sqrt{4\pi e^{2}n/\epsilon _{\infty }m^{\ast }}$, where $\epsilon
_{\infty }$ is the high-frequency contribution to the dielectric
function. In Fig.~\ref{fig:scaling} the plasmon due to heavy
quasiparticles is clearly observed at $\simeq $\thinspace
74\thinspace cm$^{-1}$ at 10\thinspace K. As temperature increases
the plasmon shifts to higher energies, implying a decrease of the
quasiparticle effective mass ($\omega _{p}^{\ast }\sim
1/\sqrt{m^{\ast }}$). Similarly, as the magnetic field increases,
the mode hardens to 90 cm$^{-1}$, which again indicates that the
heavy quasiparticles reduce their effective mass. In the top panel
of Fig.~\ref{fig:scaling}, we plot the field dependence of the
mass determined from the frequency of the peak in the loss
function spectra. In full agreement with the values obtained by
extrapolating m$^{\ast }$($\omega $) spectra to zero frequency, we
observe a marked depression of the effective mass by nearly 25$\%$
at 17\thinspace T, with the characteristic downward curvature.

We have also employed two additional methods of quantifying the
field-induced changes of $m^*$. The simplest estimate of
renormalization is the (inverse of the) position of the minimum in
reflectance. The minimum in reflectance, the so-called plasma
minimum, is determined by the screen plasma frequency, which on
the other hand is proportional to the ratio $n/m^*$, where $n$ is
the carrier density and $m^*$ their effective mass. The plasma
minimum points (from Fig.~\ref{fig:ref}) are also plotted in the
middle panel of Fig.~\ref{fig:scaling} and they agree well with
previous two data sets. Finally the partial sum rule:

\begin{equation}
\int_{0}^{\omega_c} \sigma_1(\omega) d\omega =
\frac{(\omega_p^*)^2}{8}= \frac{ \pi n e^2}{2 m^*},
\label{eq:sum}
\end{equation}
can also give an estimate of the effective mass \cite{degiorgi99}.
The upper integration limit $\omega_c$ was set just below the
onset of hybridiztion gap. The values of the effective mass
extracted this way are also in good agreement with the other three
methods (Fig.~\ref{fig:scaling}, middle panel). The scatter of
data points in Fig.~\ref{fig:scaling} serves as an estimate of the
error bars.

The Periodic Anderson Model (PAM) is believed to capture the
essence of the physics in HF metals \cite{hewson97}. We will show
below that this model when augmented with the Zeeman term
elucidates the behavior of the effective mass in a magnetic field.
At the mean-field level, the heavy fermion state is characterized
by hybridization of the free ($c$) and localized ($f$) electrons
\cite{Coleman87,Millis87}. Hybridization breaks the conduction
band into disjoint upper and lower branches with very shallow
dispersion near the gap edge (Fig.~\ref{fig:scaling} bottom
panel); this accounts for the very large effective mass of the
quasiparticles. The state is metallic and possesses a well-defined
Fermi surface that incorporates both the $c$ and $f$ electrons. In
the presence of a Zeeman term -- which consists of a magnetic
field $\vec{H}$ coupled to the total magnetic moment at each site
-- the mass enhancement factor behaves as \cite{beach04}:

\begin{equation}  \label{eq:effmass}
\biggl(\frac{m^*}{m_e}\biggr)_{\!s} =
\biggl(\frac{m^*}{m_e}\biggr)_{\!0} \biggl[ 1 +
2s\biggl(\frac{H}{H_0}\biggr) + \frac{3}{2} \biggl(\frac{H}{H_0}
\biggr)^2 \biggr],
\end{equation}
where $m_e$ is the free electron mass, $s = \pm \
(\uparrow,\downarrow)$ and $H_0$ is on the order of the Kondo
energy in magnetic units. The linear term in Eq.~\ref{eq:effmass}
results from the splitting of the Fermi surface into separate
spin-up and spin-down surfaces, each with a different band
curvature. The quadratic term is due to higher-order changes in
the hybridization energy.

The quasiparticle contribution to the optical conductivity $\sigma
(\omega )$ of the Zeeman-split heavy electron metal is:

\begin{equation}  \label{eq:conductivity}
\sigma(\omega) = \frac{e^2}{m_e} \sum_s \frac{n_s}{1/\tau -
i\omega (m^*/m_e)_s},
\end{equation}
where $1/\tau$ is the quasiparticle scattering rate. The
quasiparticles have a total density $n = n_{\uparrow} +
n_{\downarrow} = n_c + n_f$; in a Zeeman field, they are
spin-polarized according to $n_{\uparrow} - n_{\downarrow} =
H/H_0$. In the limit $\omega \tau \gg 1$, Eqs.~\ref{eq:effmass}
and \ref{eq:conductivity} can be combined to give:

\begin{equation}
-\frac{\omega _{p}^{2}}{4\pi \omega }Im\Big[\frac{1}{\sigma
(\omega )}\Big]= \biggl(\frac{m^{\ast
}}{m_e}\biggr)_{0}\biggl[1-\biggl(\frac{5}{2}-\frac{2n_{f}
}{n}\biggr)\biggl(\frac{H}{H_{0}}\biggr)^{2}\biggr],
\label{eq:field}
\end{equation}
which for H=\thinspace 0\thinspace T reduces to Eq.~\ref{eq:mass}.
According to Eq.~\ref{eq:field}, the effective mass probed in an
IR experiment is averaged between the two spin channels and its
magnitude is reduced as (H/H$_{0}$)$^{2}$, in full agreement with
the experimental observation (top panel of
Fig.~\ref{fig:scaling}). The fit (gray line) corresponds to
H$_{0}$=35\thinspace T. This estimate of $H_{0}$ is lower then the
Kondo temperature T$_{K}\simeq $\thinspace 100\thinspace K
\cite{bauer01} for CeRu$_{4}$Sb$_{12}$ but is close to the
coherence temperature T$^{\ast }\simeq 50$ K. The Zeeman splitting
also implies an overall reduction of the energy gap between the
spin-down sub-band at $E>E_{F}$ and the spin-up sub-band at
$E<E_{F}$ \cite{beach04}. As pointed out above, appreciable
reduction of the gap is not observed in our data. This is
consistent with the notion that in an optical experiment one
probes separately gaps in the spin-up and spin-down channels,
provided that spin flip scattering does not significantly alter
the selection rules. Neither of the two transitions is expected to
significantly change in fields smaller than the hybridization gap.
Further theoretical studies are needed to address the issue of
spin-flip scattering, as well as the problem of the field
dependence of the hybridization energy.

In conclusion, we have presented a comprehensive set of infrared
data obtained on CeRu$_{4}$Sb$_{12}$ in high magnetic field. The
external field is found to affect the low-energy excitations of
the system, which leads to the renormalization of the
quasiparticle effective mass. Indeed, a 17\thinspace T field was
found to diminish m$^{\ast }$ by about 25\thinspace $\%$.
Mean-field solution of the PAM with a Zeeman term offers a
quantitative account for the field dependence of the effective
mass.

The research supported by NSF.

$^{\dag}$Present Address: Department of Physics, The University of
Akron, Akron, OH 44325

\end{document}